\documentclass[twocolumn]{aastex7}

\usepackage{graphicx}
\usepackage{amsmath}
\usepackage{hyperref}
\usepackage{longtable}
\usepackage{booktabs}

\newcommand{\te}{t_{\rm E}}
\newcommand{\thetae}{\theta_{\rm E}}

\newcommand{\dl}{D_{\rm L}}

\def\e{{\rm E}}

\definecolor{brown}{rgb}{0.59, 0.29, 0.0}
\definecolor{darkgreen}{rgb}{0.0, 0.42, 0.24}
\definecolor{darkblue}{rgb}{0.01, 0.31, 0.59}
\definecolor{darkblue}{rgb}{0.0, 0.25, 0.42}
\definecolor{blue}{rgb}{0.0,0.0,1.0}
\definecolor{green}{rgb}{0.0,1.0,0.0}

\begin{document}

\title{KMT-2016-BLG-1337L: A Saturn-mass planet orbiting within a binary system of low-mass stars}
\shorttitle{KMT-2016-BLG-1337L}


\author{Cheongho Han}
\affiliation{Department of Physics, Chungbuk National University, Cheongju 28644, Republic of Korea}
\email{cheongho@astroph.chungbuk.ac.kr}
\author{Chung-Uk Lee}
\affiliation{Korea Astronomy and Space Science Institute, Daejon 34055, Republic of Korea}
\email{leecu@kasi.re.kr}
\author{Ian A. Bond}
\affiliation{Institute of Natural and Mathematical Science, Massey University, Auckland 0745, New Zealand}
\email{i.a.bond@massey.ac.nz}
\author{Andrzej Udalski} 
\affiliation{Astronomical Observatory, University of Warsaw, Al.~Ujazdowskie 4, 00-478 Warszawa, Poland}
\email{udalski@astrouw.edu.pl} 
\collaboration{5}{(Leading authors)}
\author{Michael D. Albrow}   
\affiliation{University of Canterbury, Department of Physics and Astronomy, Private Bag 4800, Christchurch 8020, New Zealand}
\email{michael.albrow@canterbury.ac.nz}
\author{Sun-Ju Chung}
\affiliation{Korea Astronomy and Space Science Institute, Daejon 34055, Republic of Korea}
\email{sjchung@kasi.re.kr}
\author{Andrew Gould}
\affiliation{Department of Astronomy, Ohio State University, 140 West 18th Ave., Columbus, OH 43210, USA}
\email{gould.34@osu.edu}
\author{Youn Kil Jung}
\affiliation{Korea Astronomy and Space Science Institute, Daejon 34055, Republic of Korea}
\affiliation{University of Science and Technology, Daejeon 34113, Republic of Korea}
\email{younkil21@gmail.com}
\author{Kyu-Ha~Hwang}
\affiliation{Korea Astronomy and Space Science Institute, Daejon 34055, Republic of Korea}
\email{kyuha@kasi.re.kr}
\author{Yoon-Hyun Ryu}
\affiliation{Korea Astronomy and Space Science Institute, Daejon 34055, Republic of Korea}
\email{yhryu@kasi.re.kr}
\author{Yossi Shvartzvald}
\affiliation{Department of Particle Physics and Astrophysics, Weizmann Institute of Science, Rehovot 76100, Israel}
\email{yossishv@gmail.com}
\author{In-Gu Shin}
\affiliation{Department of Astronomy, Westlake University, Hangzhou 310030, Zhejiang Province, China}
\email{ingushin@gmail.com}
\author{Jennifer C. Yee}
\affiliation{Center for Astrophysics $|$ Harvard \& Smithsonian 60 Garden St., Cambridge, MA 02138, USA}
\email{jyee@cfa.harvard.edu}
\author{Weicheng Zang}
\affiliation{Department of Astronomy, Westlake University, Hangzhou 310030, Zhejiang Province, China}
\email{zangweicheng@westlake.edu.cn}
\author{Hongjing Yang}
\affiliation{Department of Astronomy, Westlake University, Hangzhou 310030, Zhejiang Province, China}
\email{yanghongjing@westlake.edu.cn}
\author{Doeon Kim}
\affiliation{Department of Physics, Chungbuk National University, Cheongju 28644, Republic of Korea}
\email{qso21@hanmail.net}
\author{Dong-Jin Kim}
\affiliation{Korea Astronomy and Space Science Institute, Daejon 34055, Republic of Korea}
\email{keaton03@kasi.re.kr}
\author{Sang-Mok Cha}
\affiliation{Korea Astronomy and Space Science Institute, Daejon 34055, Republic of Korea}
\affiliation{School of Space Research, Kyung Hee University, Yongin, Kyeonggi 17104, Republic of Korea}
\email{chasm@kasi.re.kr}
\author{Seung-Lee Kim}
\affiliation{Korea Astronomy and Space Science Institute, Daejon 34055, Republic of Korea}
\email{slkim@kasi.re.kr}
\author{Dong-Joo Lee}
\affiliation{Korea Astronomy and Space Science Institute, Daejon 34055, Republic of Korea}
\email{marin678@kasi.re.kr}
\author{Yongseok Lee}
\affiliation{Korea Astronomy and Space Science Institute, Daejon 34055, Republic of Korea}
\affiliation{School of Space Research, Kyung Hee University, Yongin, Kyeonggi 17104, Republic of Korea}
\email{yslee@kasi.re.kr}
\author{Byeong-Gon Park}
\affiliation{Korea Astronomy and Space Science Institute, Daejon 34055, Republic of Korea}
\email{bgpark@kasi.re.kr}
\author{Richard W. Pogge}
\affiliation{Department of Astronomy, Ohio State University, 140 West 18th Ave., Columbus, OH 43210, USA}
\email{pogge.1@osu.edu}
\collaboration{100}{(KMTNet Collaboration)}
\author{Fumio Abe}
\affiliation{Institute for Space-Earth Environmental Research, Nagoya University, Nagoya 464-8601, Japan}
\email{abe@isee.nagoya-u.ac.jp}
\author{David P. Bennett}
\affiliation{Code 667, NASA Goddard Space Flight Center, Greenbelt, MD 20771, USA}
\affiliation{Department of Astronomy, University of Maryland, College Park, MD 20742, USA}
\email{bennett.moa@gmail.com}
\author{Aparna Bhattacharya}
\affiliation{Code 667, NASA Goddard Space Flight Center, Greenbelt, MD 20771, USA}
\affiliation{Department of Astronomy, University of Maryland, College Park, MD 20742, USA}
\email{aparna.bhattacharya@nasa.gov}
\author{Ryusei Hamada}
\affiliation{Department of Earth and Space Science, Graduate School of Science, Osaka University, Toyonaka, Osaka 560-0043, Japan}
\email{hryusei@iral.ess.sci.osaka-u.ac.jp}
\author{Yuki Hirao}
\affiliation{Institute of Astronomy, Graduate School of Science, The University of Tokyo, 2-21-1 Osawa, Mitaka, Tokyo 181-0015, Japan}
\email{hirao@ioa.s.u-tokyo.ac.jp}
\author{Asahi Idei}
\affiliation{Department of Earth and Space Science, Graduate School of Science, Osaka University, Toyonaka, Osaka 560-0043, Japan}
\email{dei@iral.ess.sci.osaka-u.ac.jp}
\author{Stela Ishitani Silva}
\affiliation{Code 667, NASA Goddard Space Flight Center, Greenbelt, MD 20771, USA}
\email{ishitanisilva@cua.edu}
\author{Shota Miyazaki}
\affiliation{Department of Earth and Space Science, Graduate School of Science, Osaka University, Toyonaka, Osaka 560-0043, Japan}
\email{miyazaki@ir.isas.jaxa.jp}
\author{Yasushi Muraki}
\affiliation{Institute for Space-Earth Environmental Research, Nagoya University, Nagoya 464-8601, Jap}
\email{muraki@isee.nagoya-u.ac.jp}
\author{Tutumi Nagai}
\affiliation{Department of Earth and Space Science, Graduate School of Science, Osaka University, Toyonaka, Osaka 560-0043, Japan}
\email{nagai@iral.ess.sci.osaka-u.ac.jp}
\author{Kansuke Nunota}
\affiliation{Department of Earth and Space Science, Graduate School of Science, Osaka University, Toyonaka, Osaka 560-0043, Japan}
\email{nunota@iral.ess.sci.osaka-u.ac.jp}
\author{Greg Olmschenk}
\affiliation{Code 667, NASA Goddard Space Flight Center, Greenbelt, MD 20771, USA}
\email{reg@olmschenk.com}
\author{Cl{\'e}ment Ranc}
\affiliation{Sorbonne Universit\'e, CNRS, UMR 7095, Institut d'Astrophysique de Paris, 98 bis bd Arago, 75014 Paris, France}
\email{ranc@iap.fr}
\author{Nicholas J. Rattenbury}
\affiliation{Department of Physics, University of Auckland, Private Bag 92019, Auckland, New Zealand}
\email{n.rattenbury@auckland.ac.nz}
\author{Yuki Satoh}
\affiliation{Department of Earth and Space Science, Graduate School of Science, Osaka University, Toyonaka, Osaka 560-0043, Japan}
\email{yukisato@kanto-gakuin.ac.jp}
\author{Takahiro Sumi}
\affiliation{Department of Earth and Space Science, Graduate School of Science, Osaka University, Toyonaka, Osaka 560-0043, Japan}
\email{sumi@ess.sci.osaka-u.ac.jp}
\author{Daisuke Suzuki}
\affiliation{Department of Earth and Space Science, Graduate School of Science, Osaka University, Toyonaka, Osaka 560-0043, Japan}
\email{dsuzuki@ir.isas.jaxa.jp}
\author{Takuto Tamaoki}
\affiliation{Department of Earth and Space Science, Graduate School of Science, Osaka University, Toyonaka, Osaka 560-0043, Japan}
\email{tamaoki@iral.ess.sci.osaka-u.ac.jp}
\author{Sean K. Terry}
\affiliation{Code 667, NASA Goddard Space Flight Center, Greenbelt, MD 20771, USA}
\affiliation{Department of Astronomy, University of Maryland, College Park, MD 20742, USA}
\email{skterry@umd.edu}
\author{Paul J. Tristram}
\affiliation{University of Canterbury Mt.~John Observatory, P.O. Box 56, Lake Tekapo 8770, New Zealand }
\email{tristram.p@gmail.com}
\author{Aikaterini Vandorou}
\affiliation{Code 667, NASA Goddard Space Flight Center, Greenbelt, MD 20771, USA}
\affiliation{Department of Astronomy, University of Maryland, College Park, MD 20742, USA}
\email{aikaterini.vandorou@utas.edu.au}
\author{Hibiki Yama}
\affiliation{Department of Earth and Space Science, Graduate School of Science, Osaka University, Toyonaka, Osaka 560-0043, Japan}
\email{yama@iral.ess.sci.osaka-u.ac.jp}
\collaboration{100}{(MOA Collaboration)}
\author{Przemek Mr{\'o}z}
\affiliation{Astronomical Observatory, University of Warsaw, Al.~Ujazdowskie 4, 00-478 Warszawa, Poland}
\email{pmroz@astrouw.edu.pl}
\author{Micha{\l} K. Szyma{\'n}ski}
\affiliation{Astronomical Observatory, University of Warsaw, Al.~Ujazdowskie 4, 00-478 Warszawa, Poland}
\email{msz@astrouw.edu.pl}
\author{Jan Skowron}
\affiliation{Astronomical Observatory, University of Warsaw, Al.~Ujazdowskie 4, 00-478 Warszawa, Poland}
\email{jskowron@astrouw.edu.pl}
\author{Rados{\l}aw Poleski} 
\affiliation{Astronomical Observatory, University of Warsaw, Al.~Ujazdowskie 4, 00-478 Warszawa, Poland}
\email{radek.poleski@gmail.co}
\author{Igor Soszy{\'n}ski}
\affiliation{Astronomical Observatory, University of Warsaw, Al.~Ujazdowskie 4, 00-478 Warszawa, Poland}
\email{soszynsk@astrouw.edu.pl}
\author{Pawe{\l} Pietrukowicz}
\affiliation{Astronomical Observatory, University of Warsaw, Al.~Ujazdowskie 4, 00-478 Warszawa, Poland}
\email{pietruk@astrouw.edu.pl}
\author{Szymon Koz{\l}owski} 
\affiliation{Astronomical Observatory, University of Warsaw, Al.~Ujazdowskie 4, 00-478 Warszawa, Poland}
\email{simkoz@astrouw.edu.pl}
\author{Krzysztof A. Rybicki}
\affiliation{Astronomical Observatory, University of Warsaw, Al.~Ujazdowskie 4, 00-478 Warszawa, Poland}
\affiliation{Department of Particle Physics and Astrophysics, Weizmann Institute of Science, Rehovot 76100, Israel}
\email{krybicki@astrouw.edu.pl}
\author{Patryk Iwanek}
\affiliation{Astronomical Observatory, University of Warsaw, Al.~Ujazdowskie 4, 00-478 Warszawa, Poland}
\email{piwanek@astrouw.edu.pl}
\author{Krzysztof Ulaczyk}
\affiliation{Department of Physics, University of Warwick, Gibbet Hill Road, Coventry, CV4 7AL, UK}
\email{kulaczyk@astrouw.edu.pl}
\author{Marcin Wrona}
\affiliation{Astronomical Observatory, University of Warsaw, Al.~Ujazdowskie 4, 00-478 Warszawa, Poland}
\affiliation{Villanova University, Department of Astrophysics and Planetary Sciences, 800 Lancaster Ave., Villanova, PA 19085, USA}
\email{mwrona@astrouw.edu.pl}
\author{Mariusz Gromadzki}          
\affiliation{Astronomical Observatory, University of Warsaw, Al.~Ujazdowskie 4, 00-478 Warszawa, Poland}
\email{marg@astrouw.edu.pl}
\author{Mateusz J. Mr{\'o}z} 
\affiliation{Astronomical Observatory, University of Warsaw, Al.~Ujazdowskie 4, 00-478 Warszawa, Poland}
\email{mmroz@astrouw.edu.pl}
\collaboration{100}{(The OGLE Collaboration)}

\correspondingauthor{\texttt{leecu@kasi.re.kr}}

\begin{abstract}
We report the discovery and characterization of a planetary companion in the 
microlensing event KMT-2016-BLG-1337, which was produced by a binary system 
of low-mass stars. The light curve of the event exhibits a short-term anomaly 
superposed on the profile of a binary-lens single-source (2L1S) model.  To 
investigate the nature of this anomaly, we performed detailed modeling under 
both the binary-lens binary-source (2L2S) and triple-lens single-source (3L1S) 
interpretations. The 3L1S model provides a substantially better fit to the data, 
strongly favoring the presence of a planetary companion in the lens system.  Two viable 
$3L1S$ solutions describe the event nearly equally well. In one solution, the planet 
has a mass of $M_3 \sim 0.3~M_{\mathrm{J}}$ and lies at a projected separation of 
$a_{\perp,3} \sim 4~{\rm au}$ from the heavier member of the host binary.  In the 
alternative solution, the planet has a mass of $M_3 \sim 7~M_{\mathrm{J}}$ and a 
projected separation of $a_{\perp,3} \sim 1.5~{\rm au}$.  The host binary consists 
of early M-type dwarfs with masses of $M_1 \sim 0.54~M_\odot$ and 
$M_2 \sim 0.40~M_\odot$, separated in projection by $a_{\perp,2} \sim 3.5~{\rm au}$.  
The system is located at a distance of $D_{\rm L} \sim 7~{\rm kpc}$ toward the Galactic 
bulge.  This event demonstrates the sensitivity of microlensing to planets in dynamically 
complex stellar environments, including systems beyond the reach of other detection 
techniques.  It thereby contributes to a more comprehensive understanding of planet 
formation in multiple-star systems.
\end{abstract}

\keywords{Gravitational microlensing (672) -- exoplanet detection (2147)}

\section{Introduction} \label{sec:one}

Detecting planets in binary star systems is crucial for deepening our 
understanding of how planets form and evolve. Traditional planet formation 
theories were developed primarily for single-star systems, like our Solar System, 
but binary systems introduce additional gravitational complexities that can 
disrupt or reshape protoplanetary disks. The discovery of planets within these 
systems challenges classical assumptions and demonstrates that planet formation 
is more resilient and adaptable than once believed. Such findings help refine 
theoretical models such as core accretion and disk instability, leading to a 
more comprehensive view of planetary origins.

The study of planets in binaries also holds significant statistical importance. 
Since roughly half of all stars in the Milky Way exist in binary or multiple 
systems, excluding them would leave our understanding of planetary populations 
incomplete.  By including binary systems in exoplanet surveys, astronomers can 
develop a more accurate estimate of how common planetary systems are throughout 
the Galaxy and how stellar multiplicity influences their distribution.

The microlensing technique plays a uniquely powerful role in this field. It can 
reveal planets found in many types of binary systems, including those that orbit 
both stars (circumbinary systems), those that orbit only one star in the pair
\citep{Han2017a},  
and those that belong to one star while the other influences the gravitational 
lensing effect. This flexibility makes microlensing more effective for studying 
complex systems than most other planet detection methods.

\begin{figure*}[t]
\centering
\includegraphics[width=13.5cm]{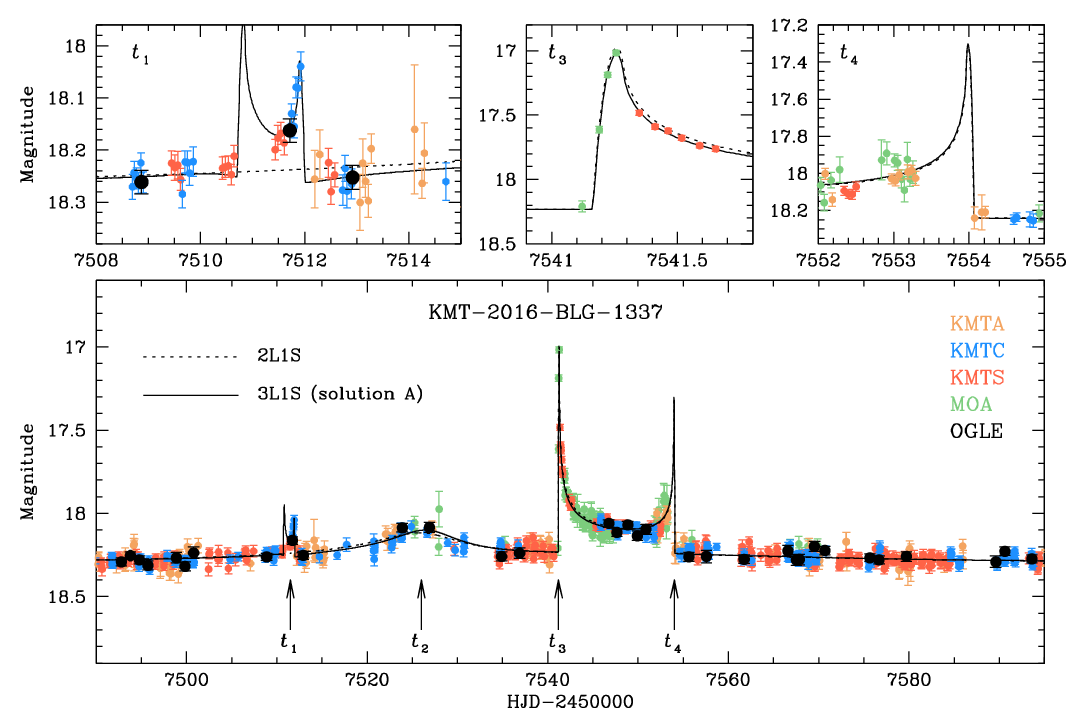}
\caption{
Light curve of the microlensing event KMT-2016-BLG-1337. The lower panel presents 
the full light curve of the event, with labeled arrows marking the times of the 
major anomaly features. The upper panels provide enlarged views of the regions 
around $t_1$, $t_3$, and $t_4$.  The dotted and solid curves plotted over the 
data points represent the model fits for the binary-lens single-source (2L1S) and 
triple-lens single-source (3L1S: solution A) models, respectively.  The colors of 
the data points correspond to the telescopes used for the observations, as indicated 
in the legend.
}
\label{fig:one}
\end{figure*}

Another key advantage of microlensing is its independence from the light of the 
host star.  Because it relies on gravitational effects rather than direct light 
detection, microlensing can identify planets around faint or non-luminous stars, 
such as M-dwarfs or brown dwarfs, even when the system cannot be resolved by 
telescopes. Moreover, microlensing allows astronomers to detect planets in distant 
regions of the Galaxy, far beyond the reach of other detection techniques, helping 
build a more complete Galactic census of planetary systems.

Microlensing surveys conducted since the early 1990s have led to the discovery of 
ten  planets orbiting within binary systems.  The lens system OGLE-2006-BLG-284L 
consists of two low-mass stars and a gas giant planet with a mass ratio of $q=1.26 
\times 10^{-3}$ to the primary \citep{Bennett2020}.  The event OGLE-2013-BLG-0341 
revealed a cold super-Earth orbiting a low-mass star in a binary separated by about 
15 au, located beyond the system’s snow line \citep{Gould2014}. OGLE-2008-BLG-092 
was modeled as a triple-lens system comprising two stars and an ice giant similar 
to Uranus, showing that giant planets can form in dynamically perturbed binaries 
\citep{Poleski2014}. OGLE-2007-BLG-349L marked the first discovery of a circumbinary 
planet through microlensing, revealing a Saturn-mass planet orbiting both stars in 
the system \citep{Bennett2016}. In OGLE-2016-BLG-0613, the lens model involved a 
super-Jupiter bound to one star in a binary, producing complex caustic patterns from 
triple-lens interactions \citep{Han2017b}. OGLE-2018-BLG-1700 revealed a planet that 
could be either circumprimary or circumbinary, highlighting the degeneracies in 
multi-lens modeling \citep{Han2020}.  In the case of the lensing event KMT-2019-BLG-1715, 
the lens consists of planet-mass object with $\sim 2.6~M_{\rm J}$ and binary stars of 
K and M dwarfs lying in the galactic disk \citep{Han2021} KMT-2020-BLG-0414 is an 
exceptional case in which an Earth-mass planet resides in a binary composed of an 
M dwarf and a low-mass brown dwarf \citep{Zang2021a}.   The event OGLE-2023-BLG-0836 
required a triple-mass lens model, identifying a binary system with a planetary 
companion and marking the sixth such discovery \citep{Han2024}. Finally, 
KMT-2024-BLG-0404L involved a star, brown dwarf, and planet, extending microlensing 
discoveries to mixed stellar--substellar binaries \citep{Han2025b}. Collectively, 
these detections demonstrate that planet formation and survival are feasible across 
a wide range of binary configurations, underscoring the resilience and adaptability 
of planetary systems in diverse dynamical environments.

In this work, we report the discovery of a new microlensing planet residing in a binary 
stellar system. This planet was identified through a detailed reanalysis of observational 
data obtained from lensing surveys. The detection adds to the gradually increasing 
number of microlensing planets found in binary configurations, further illustrating 
the capability of microlensing to uncover planetary systems that are otherwise 
difficult to detect. By expanding the sample of known planets in binary systems, 
this discovery provides an additional data point for testing models of planet 
formation and stability under the complex gravitational conditions of multiple-star 
environments.

\begin{figure*}[t]
\centering
\includegraphics[width=16.5cm]{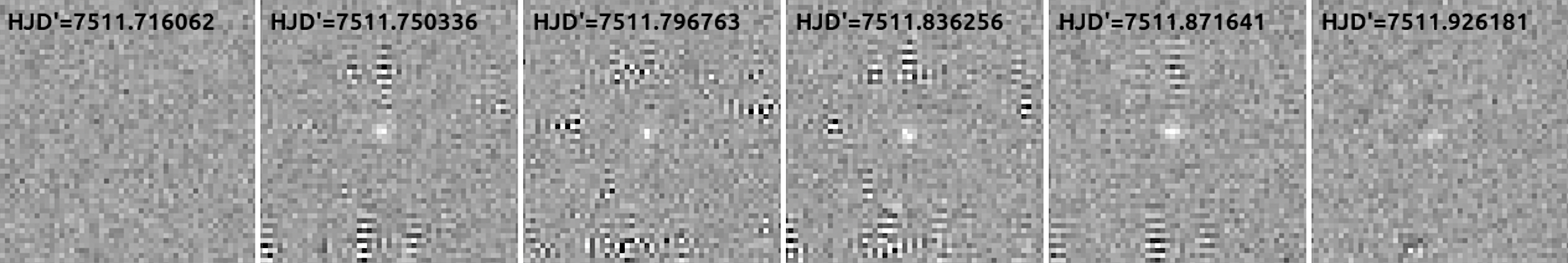}
\caption{
Difference images taken during the planet-induced anomaly around $t_1$. The label 
in each panel indicates the time at which the image was obtained. 
}
\label{fig:two}
\end{figure*}

\section{Identification of planetary signal} \label{sec:two}

The newly discovered planet was identified through a reanalysis of microlensing 
events detected by the Korea Microlensing Telescope Network (KMTNet) survey 
\citep{Kim2016}, which has been in operation since 2015. The primary objective 
of this reanalysis was to search for subtle planetary signals that may have been
missed in previously modeled binary-lens (2L1S) events.

Each year, during the observing season from March to October, the KMTNet survey 
detects approximately 3,000 microlensing events. The majority of these events are 
caused by single-lens single-source (1L1S) configurations, for which lensing models 
are automatically generated in near real-time upon event detection \citep{Kim2018}.  
Planetary signals in these events are initially identified through visual inspection 
of the resulting light curves. In addition, in order to perform a uniform, objective, 
and automated search for anomalies across the entire dataset, the KMTNet collaboration 
employs an automated algorithm. This algorithm, AnomalyFinder \citep{Zang2021b}, 
identifies potential planetary signals by analyzing the residuals between observed 
data and the best-fit 1L1S microlensing model.

In the case of binary-lens events, short-duration planetary signals can occur when 
a planetary companion perturbs the magnification pattern produced by the binary-lens 
system \citep{Lee2008}.  However, detecting such planetary signals using algorithms 
like AnomalyFinder requires an analysis of the residuals from the binary-lens (2L1S) 
model. Currently, the modeling of 2L1S events has not yet been fully automated, and 
therefore, in order to identify planetary signals in these cases, a manual modeling 
of the 2L1S events must first be conducted. The residuals from these models are then 
subject to visual inspection to detect possible planetary anomalies. The newly 
discovered planet orbiting within a binary system, KMT-2016-BLG-1337Lb, was uncovered 
through a reanalysis of previously detected binary-lens microlensing events from the 
KMTNet survey.

Figure~\ref{fig:one} presents the light curve of the lensing event KMT-2016-BLG-1337, 
derived from observations obtained by the three KMTNet telescopes and supplemented 
with data from the Microlensing Observations in Astrophysics (MOA) survey \citep{Bond2001, 
Sumi2003} and the Optical Gravitational Lensing Experiment (OGLE) survey 
\citep[OGLE;][]{Udalski2015}.  The light curve displays two distinct spikes at approximately 
$t_3 \sim 7541.2$ and $t_4 \sim 7554.0$, together with a broad, weak bump centered around 
$t_2 \sim 7526.0$.  This overall pattern is characteristic of a caustic-crossing 2L1S event, 
in which the two spikes likely correspond to the source’s caustic crossings, and the weak 
bump is probably caused by the source’s approach to a cusp of the caustic.

However, a careful inspection of the data revealed a short-term anomaly that could not 
be explained by the 2L1S model. The anomaly occurred around $t_1 \sim 7511$ and lasted 
for approximately one day. It shows a characteristic pattern produced by the source’s 
crossing of a tiny caustic, with the second caustic crossing clearly resolved in the data. 
The anomaly was independently detected in two KMTNet datasets (KMTC and KMTS). In addition, 
a single data point from the OGLE survey corroborates the anomaly. These independent 
detections strongly suggest that the anomaly represents a genuine astrophysical signal.
This interpretation is further supported by the inspection of the images obtained during 
the anomaly, as shown in Figure~\ref{fig:two} . In the following sections, we describe 
the data acquisition process (Section~\ref{sec:three}) and provide a detailed analysis 
of the event (Section~\ref{sec:four}).

\begin{deluxetable*}{lllllllll}
\tablewidth{0pt}
\tablecaption{Lensing parameters of 2L1S, 2L2S, and 3L1S solutions. \label{table:one}}
\tablehead{
\multicolumn{1}{c}{Parameter}   &
\multicolumn{1}{c}{2L1S}        &
\multicolumn{1}{c}{2L2S}        &
\multicolumn{2}{c}{3L1S}        \\
\multicolumn{1}{c}{}            &
\multicolumn{1}{c}{}            &
\multicolumn{1}{c}{}            &
\multicolumn{1}{c}{Solution A}  &
\multicolumn{1}{c}{Solution B}  
}
\startdata
$\chi^2$                    &  $1719.1           $         &   $1531.6           $   &  $1487.5           $    &  $1490.2           $ \\
$t_0$ (HJD$^\prime$)        &  $7538.04 \pm 0.12 $         &   $7540.35 \pm 0.40 $   &  $7537.99 \pm 0.44 $    &  $7538.34 \pm 0.40 $ \\
$u_0$                       &  $0.2703 \pm 0.0050$         &   $0.2227 \pm 0.0050$   &  $0.2959 \pm 0.0056$    &  $0.2989 \pm 0.0074$ \\
$\te$ (days)                &  $47.12 \pm 0.98   $         &   $52.33 \pm 1.06   $   &  $41.10 \pm 1.10   $    &  $40.07 \pm 1.14   $ \\
$s_2$                       &  $0.9376 \pm 0.0049$         &   $0.8884 \pm 0.0056$   &  $0.9582 \pm 0.0041$    &  $0.9620 \pm 0.0048$ \\
$q_2$                       &  $0.658 \pm 0.041  $         &   $0.812 \pm 0.050  $   &  $0.739 \pm 0.052  $    &  $0.819 \pm 0.032  $ \\
$\alpha$ (rad)              &  $2.4675 \pm 0.0076$         &   $2.5315 \pm 0.0200$   &  $2.4535 \pm 0.0221$    &  $2.4647 \pm 0.0192$ \\
$s_3$                       &   \nodata                    &    \nodata              &  $1.0741 \pm 0.0085$    &  $0.3939 \pm 0.0063$ \\
$q_3$ ($10^{-3}$)           &   \nodata                    &    \nodata              &  $0.49 \pm 0.11    $    &  $12.55 \pm 2.34   $ \\
$\psi$ (rad)                &   \nodata                    &    \nodata              &  $3.591 \pm 0.032  $    &  $0.336 \pm 0.020  $ \\
$\rho$ ($10^{-3}$)          &  $0.962 \pm 0.51   $         &   $0.74 \pm 0.04    $   &  $1.076 \pm 0.057  $    &  $1.082 \pm 0.061  $ \\
$t_{0,2}$ (HJD$^\prime$)    &   \nodata                    &   $7530.33 \pm 0.97 $   &   \nodata               &   \nodata            \\
$u_{0,2}$                   &   \nodata                    &   -$0.694 \pm 0.016 $   &   \nodata               &   \nodata            \\
$\rho_2$ ($10^{-3}$)        &   \nodata                    &   \nodata               &   \nodata               &   \nodata            \\
$q_F$                       &   \nodata                    &   $0.222 \pm 0.028  $   &   \nodata               &   \nodata            \\
\enddata
\tablecomments{
${\rm HJD}^\prime\equiv {\rm HJD}-2450000$.
}
\end{deluxetable*}

\section{Observations and data} \label{sec:three}

The lensing event KMT-2016-BLG-1337 was identified by the KMTNet survey from observations 
of a source star located toward the Galactic bulge. The equatorial coordinates of the 
source are $(\mathrm{R.A.}, \mathrm{Decl.}) = (17{:}56{:}31.21, -33{:}02{:}03.08)$, 
corresponding to the Galactic coordinates $(l, b) = (-2^\circ\hskip-2pt .3340, 
-4^\circ\hskip-2pt .0998)$.  The source has a baseline magnitude of $I_{\rm base} = 18.62$, 
and the $I$-band extinction toward the field is $A_I = 1.23$.  The event was also independently 
detected by the MOA survey and designated MOA-BLG-2016-292.  Although it was not initially 
identified by the OGLE survey because of the relatively sparse coverage of this region, OGLE 
data were subsequently recovered through additional photometric analysis of the source 
identified by the other surveys. In this paper, we refer to the event using the KMTNet 
designation.

Observations of the event were conducted using telescopes operated by the three survey 
groups.  The KMTNet survey employs a system of three identical 1.6-meter telescopes, 
strategically located to provide nearly continuous time coverage of the same sky fields 
as the Earth rotates. The telescopes are situated at the Cerro Tololo Inter-American 
Observatory in Chile (KMTC), the South African Astronomical Observatory in South Africa 
(KMTS), and the Siding Spring Observatory in Australia (KMTA). Each KMTNet telescope is 
equipped with a mosaic CCD camera composed of four 9k × 9k detectors, offering a wide 
field of view of approximately 4 square degrees. The MOA survey utilizes a 1.8-meter 
telescope located at the Mount John University Observatory in New Zealand.  Its camera 
consists of ten 2k × 4k CCDs, providing an effective sky coverage of about 2.2 square 
degrees.  The OGLE survey operates a 1.3-m telescope at Las Campanas Observatory in Chile, 
equipped with a camera providing a 1.4 square-degree field of view.

The event was observed in the $I$ band by the KMTNet and OGLE surveys and in the customized 
MOA-$R$ band by the MOA survey. Image reduction and photometric measurements were performed 
using the respective pipelines developed by \citet{Albrow2009} for the KMTNet data, A. 
\citet{Udalski2003} for the OGLE survey, and \citet{Bond2001} for the MOA data.  These 
pipelines employ the difference image analysis technique \citep{Tomaney1996, Alard1998, 
Wozniak2000} to obtain precise photometry in crowded stellar fields. The KMTNet data were 
subsequently reprocessed using the photometry code developed by \citet{Yang2024} to ensure 
optimal photometric precision.

\section{Interpretation of the lensing event}  \label{sec:four}

To interpret the anomalous features observed in the lensing light curve, we carried 
out a series of modeling analyses under various configurations of the lens system. 
First, considering the presence of caustic spikes in the light curve, we initially 
modeled the event using a 2L1S configuration.  Subsequently, we examined the residuals 
from the 2L1S model and performed additional modeling to account for anomaly features 
that could not be explained by this configuration. In this process, we considered both 
a triple-lens single-source (3L1S) model, which introduces an additional component to 
the lens system, and a binary-lens binary-source (2L2S) model, in which the source 
consists of two stars. The results obtained from each model are presented in the 
following subsections.

\subsection{Binary-lens single-source (2L1S) model} \label{sec:four-one}

We began modeling the light curve using the 2L1S configuration. The light curve of 
a 2L1S event is described by seven fundamental parameters. Among these, the first 
three parameters $(t_0, u_0, \te)$ characterize the motion of the source relative 
to the lens. The parameter $t_0$ denotes the time of the closest approach, $u_0$ 
represents the lens-source separation at that time, and $\te$ indicates the event 
timescale. Two additional parameters $(s, q)$ define the properties of the binary 
lens, where $s$ is the projected separation between the two lens components and $q$ 
is the mass ratio of the components. The parameter $\alpha$ specifies the angle between 
the source trajectory and the binary lens axis (source incidence angle). Here, both 
$u_0$ and $s$ are expressed in units of the angular Einstein radius $\theta_{\rm E}$.  
For KMT-2016-BLG-1337, the light curve exhibits distinct caustic crossing features, 
during which finite source effects become significant.
To model these parts accurately, we introduced an additional parameter 
$\rho$ (normalized source radius), defined as the ratio of the angular source radius 
$\theta_*$ to the angular Einstein radius $\theta_{\rm E}$.

The modeling was performed to determine the set of lensing parameters, or the lensing 
solution, that best reproduces the observed light curve. In this procedure, the binary 
parameters $s$ and $q$ were explored using a grid search, while the remaining parameters 
were optimized through a downhill minimization method based on the Markov Chain Monte 
Carlo (MCMC) algorithm. The initial values of the downhill parameters were assigned 
based on the time of peak, peak magnification, and event duration. For the source 
trajectory angle, multiple initial values were evenly distributed within the range of 
$(0, 2\pi)$.

\begin{figure}[t]
\includegraphics[width=\columnwidth]{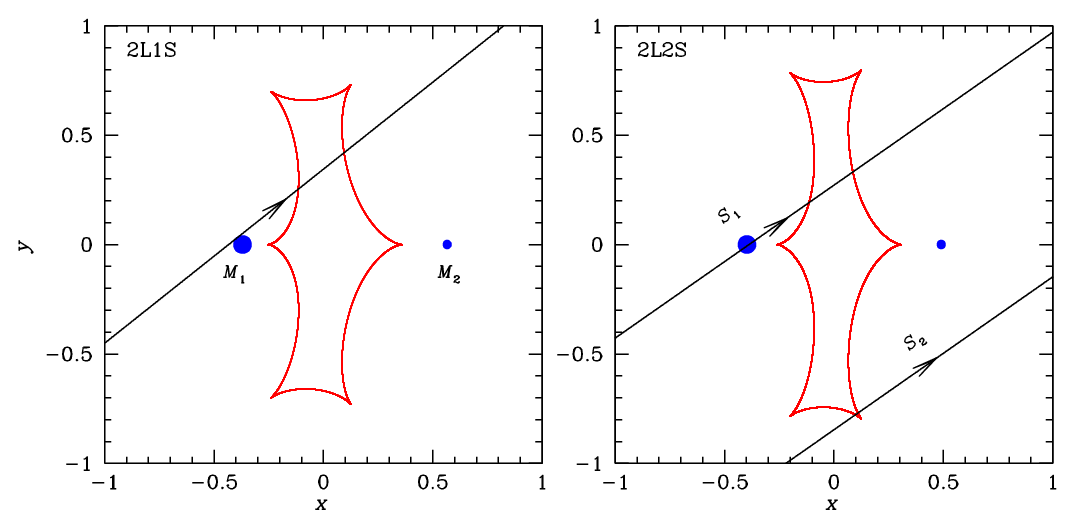}
\caption{
Lens-system configurations of the 2L1S and 2L2S models. In each panel, the two blue 
dots mark the positions of the binary lens components $M_1$ and $M_2$, where the bigger 
dot indicates the heavier component.  The red closed cuspy curve represents the caustic, 
and the arrowed line indicates the source trajectory.  In the case of the 2L2S model, 
two source trajectories are shown, one corresponding to the primary source $S_1$ and the 
other to the secondary source $S_2$.  The origin of the coordinate system is set at the 
barycenter of the lens, and spatial scales are given in units of the angular Einstein 
radius associated with the combined mass of the lens.
}
\label{fig:three}
\end{figure}

From the 2L1S modeling, we identified a solution that describes the overall pattern 
of the light curve.  The best-fit binary parameters are $(s, q) \sim (0.94, 0.66)$, 
with an event timescale of $\te \sim 47$ days. The complete set of lensing parameters 
and their uncertainties are listed in Table~\ref{table:one}.  The model curve corresponding 
to this best-fit solution is shown as a dotted curve overlaid on the observational data 
points in Figure~\ref{fig:one}.  The lens system configuration, showing the source 
trajectory relative to the lens and caustic, is presented in the left panel of 
Figure~\ref{fig:three}. It shows that the binary lens forms a resonant caustic due to the 
binary separation being close to unity, and the source trajectory passes diagonally through 
the upper region of the caustic. According to this solution, the weak feature observed 
around $t_2$ arises from the source’s approach to the left on-axis cusp of the caustic, 
while the two subsequent spikes at $t_3$ and $t_4$ are produced as the source crosses the 
caustic folds. However, the model fails to explain the short-term anomaly at around $t_1$.

\subsection{Binary-lens binary-source (2L2S) model} \label{sec:four-two}

An additional short-term anomaly observed in the light curve of a binary-lens event can 
occur when the source itself is a binary system, as demonstrated by the event KMT-2019-BLG-1715 
\citep{Han2021}. To explore this possibility, we performed a 2L2S modeling analysis to 
determine whether the anomaly observed at $t_1$, which could not be explained by the 
2L1S model, might originate from a faint companion to the source.

The 2L2S configuration corresponds to the case in which an additional source star is included
compared to the 2L1S configuration. Consequently, 2L2S modeling requires additional parameters
to describe the secondary source star ($S_2$). These parameters are $(t_{0,2}, u_{0,2}, \rho_2,
q_F)$, where the first two denote the time and separation at the closest approach, the third
represents the normalized source radius of $S_2$, and the last ($q_F$) is the flux ratio between
the secondary and primary source star ($S_1$). Because the 2L1S model adequately explains the
overall light curve, including the three anomaly features at $t_2$, $t_3$, and $t_4$, the 2L2S
modeling was performed with a focus on identifying a secondary source trajectory that could
account for the additional anomaly observed at $t_1$.

\begin{figure}[t]
\includegraphics[width=\columnwidth]{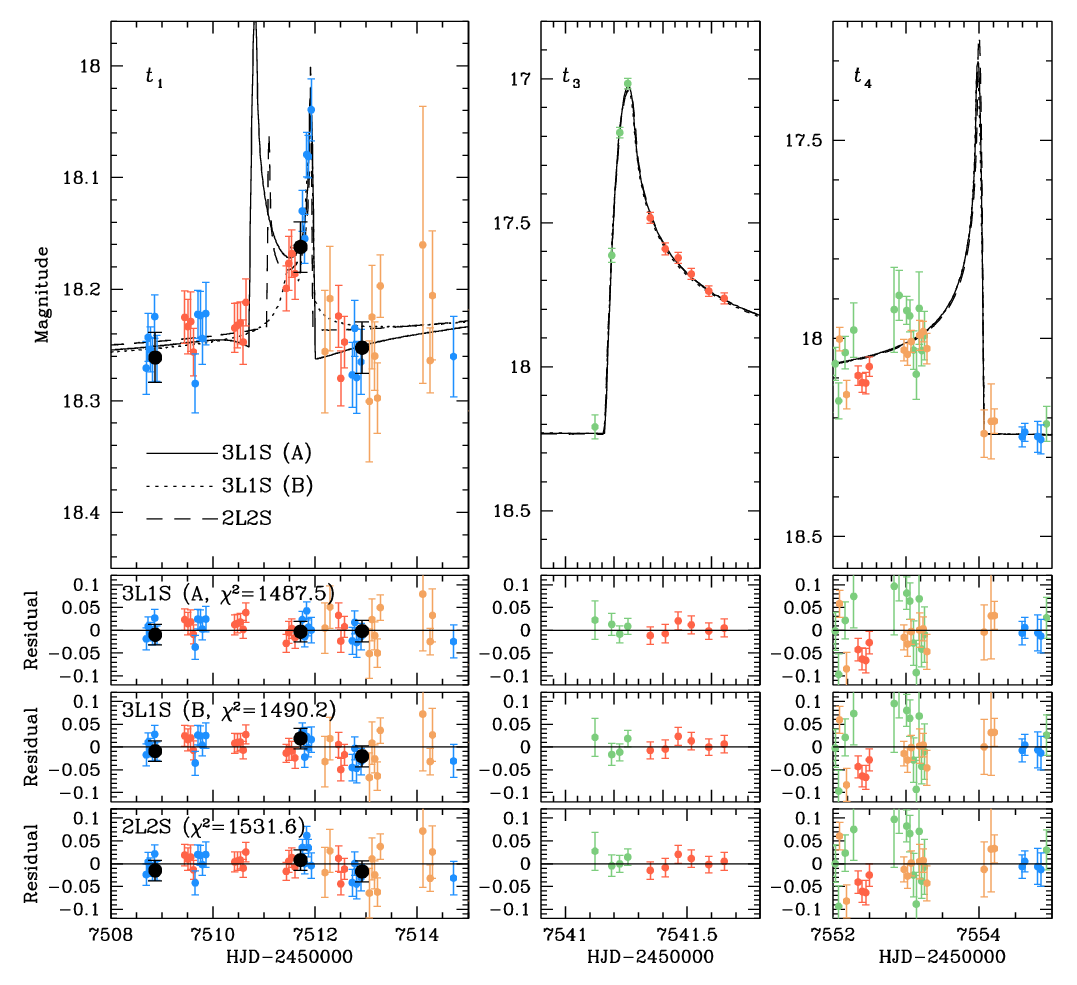}
\caption{
Comparison of the two 3L1S models (solutions A and B) and the 2L2S
model in the regions encompassing the three major anomalies around
$t_1$, $t_3$, and $t_4$.
}
\label{fig:four}
\end{figure}

In Table~\ref{table:one}, we present the lensing parameters of the best-fit 2L2S solution. 
A zoomed-in view of the model light curve in the regions around the major anomalies is shown 
in Figure~\ref{fig:four}, where the model curve (dashed line) is overlaid on the data points. 
The model reproduces the anomaly near $t_1$, as well as those around the other anomaly features, 
reasonably well. The right panel of Figure~\ref{fig:three} illustrates the corresponding lens 
configuration, indicating that the anomaly at $t_1$ arises from the secondary source passing 
over the tip of the lower-right caustic cusp. The secondary source is considerably fainter 
than the primary, with a flux ratio of $q_F \sim 0.22$.

\begin{figure}[t]
\includegraphics[width=\columnwidth]{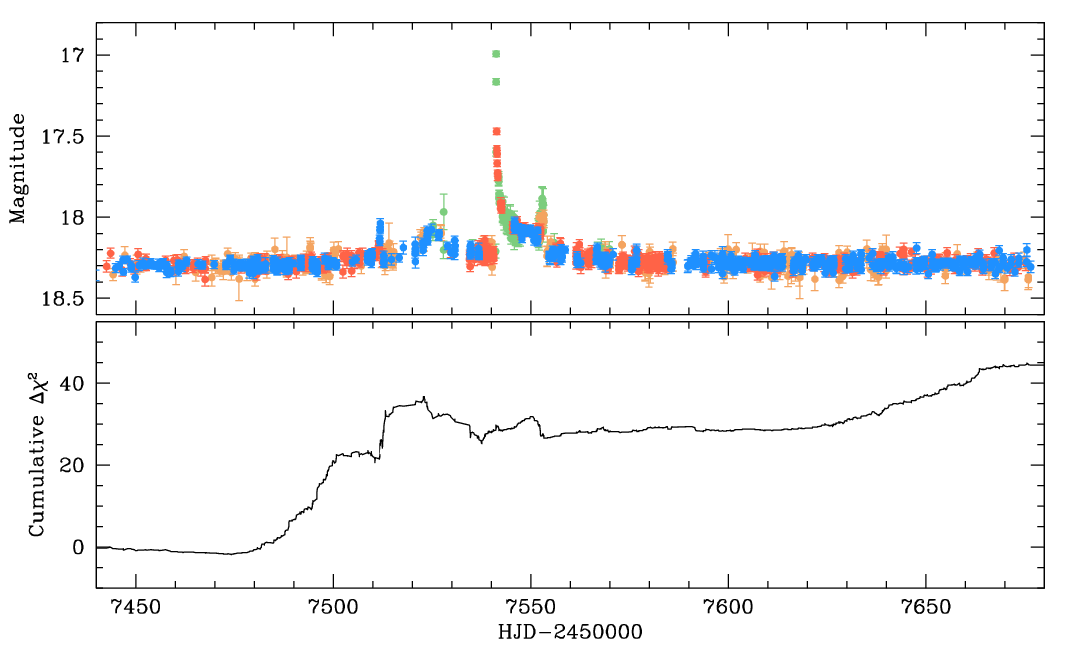}
\caption{
Cumulative distribution of $\chi^2$ difference between the 2L2S and 3L1S model. 
The light curve in the upper panel is presented to show the region of $\chi^2$ 
difference.
}
\label{fig:five}
\end{figure}

\subsection{Triple-lens single-source (3L1S) model} \label{sec:four-three}

We further explored the possibility that the anomaly observed near epoch $t_1$ was caused 
by a planet orbiting within a binary stellar system. To model this scenario, we employed 
a triple-lens framework, introducing additional parameters to describe the third (planetary) 
body. These parameters, $(s_3, q_3, \psi)$, denote the projected separation and mass ratio 
of the third body ($M_3$) relative to the primary lens ($M_1$), and the orientation angle 
($\psi$) of $M_3$ measured counterclockwise from the $M_1$--$M_2$ axis.  To avoid confusion 
with the binary-lens parameters, we denote the parameters describing the $M_1$--$M_2$ pair 
as $(s_2, q_2)$.  A comprehensive summary of the lensing parameters for the various lens-system 
configurations is provided in Table 2 of \citet{Han2023}.

The 3L1S modeling was carried out as follows. In the first step, we performed a grid search 
over the third-body parameters $(s_3, q_3, \psi)$ while fixing all other parameters to those 
derived from the best-fit 2L1S model. This strategy was motivated by the fact that the 2L1S 
model adequately reproduces the overall light curve except for the short-term anomaly around 
$t_1$. In the second step, each local solution identified from the grid search was refined 
by allowing all parameters to vary freely.

The 3L1S modeling yielded two local solutions.  The first solution (hereafter A) corresponds 
to $(s_3, q_3, \psi) \simeq (1.07, 0.49\times10^{-3}, 3.59)$, while the second (B) corresponds 
to $(s_3, q_3, \psi) \simeq (0.39, 12.6\times10^{-3}, 0.34)$. In both cases, the small mass 
ratios indicate that the third body is of planetary mass.  The 3L1S model provides a substantially 
better fit than the 2L2S model, with an improvement of $\Delta\chi^2 = 44.1$, implying that 
the lens system is a hierarchical triple composed of a planet orbiting a binary stellar host.
Figure~\ref{fig:five} shows the cumulative distribution of the $\chi^2$ difference, 
$\Delta\chi^2=\chi^2_{\rm 2L2S}-\chi^2_{\rm 3L1S}$, between the 2L2S and 3L1S models. 
The distribution shows that the major contribution to $\Delta\chi^2$ arises from the 
region around $t_1$.

The complete sets of lensing parameters for the two 3L1S solutions are listed in 
Table~\ref{table:one}.  The corresponding model light curves and residuals around the major 
anomalies are presented in Figure~\ref{fig:four}, while the full model light curve for 
solution~A is shown in Figure~\ref{fig:one}. The two solutions are nearly degenerate, with 
solution A being slightly preferred by $\Delta\chi^2 = 2.7$. A comparison of the model light 
curves reveals that solution A features two distinct caustic-crossing spikes, whereas solution 
B produces a single peak near the caustic exit. This degeneracy could have been resolved if 
the caustic entrance had been observed. Unfortunately, observations from the KMTC site during 
the critical epoch of divergence between the two models (HJD$^\prime \sim 7511$) were lost 
due to cloudy weather conditions.

\begin{figure}[t]
\includegraphics[width=\columnwidth]{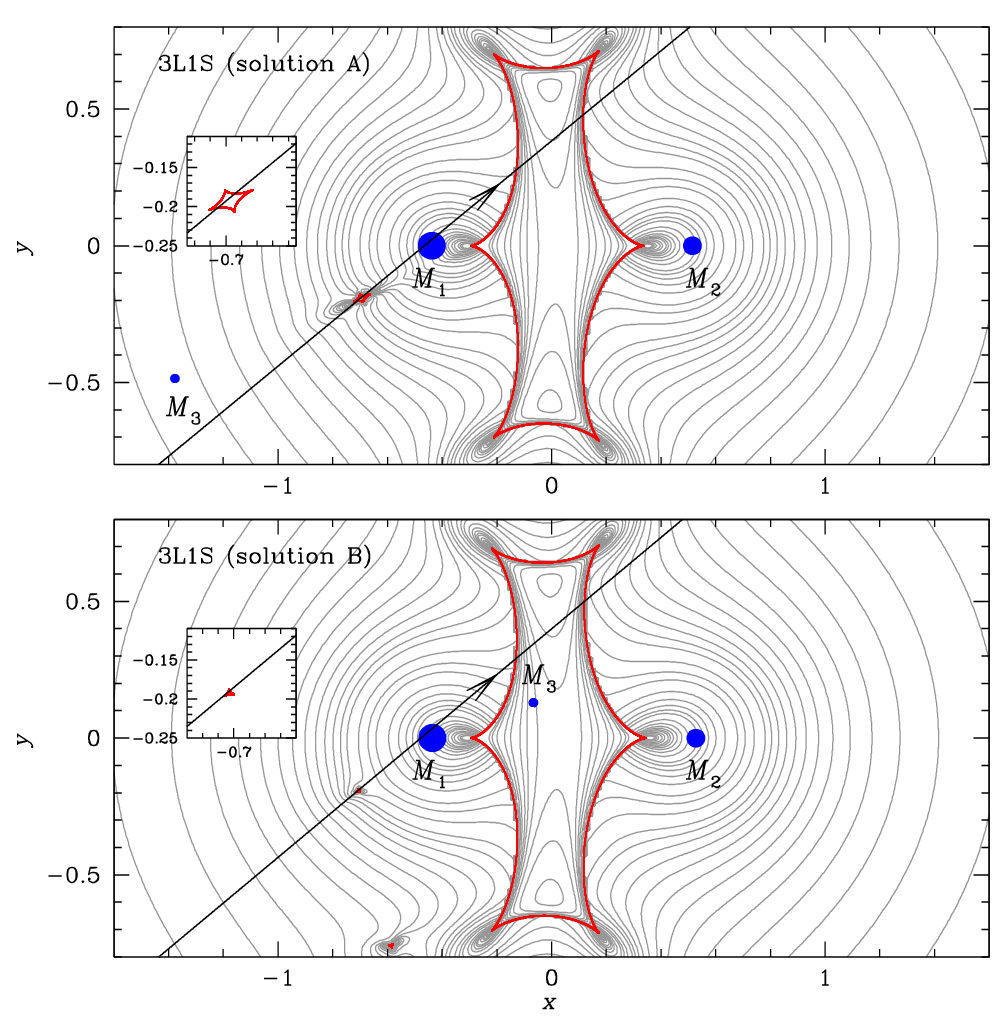}
\caption{
Lens-system configurations for the two 3L1S solutions. The gray curves surrounding the 
caustics represent contours of equal magnification. The inset in each panel presents a 
zoomed-in view of the caustic structure generated by the planetary-mass third body ($M_3$).
}
\label{fig:six}
\end{figure}

Figure~\ref{fig:six} displays the lens system configurations for solutions A (upper panel) 
and B (lower panel).  In both cases, the caustic produced by the $M_1$--$M_2$ binary is nearly 
identical to that of the 2L1S model. The planetary mass third body introduces an additional 
small caustic, and the source trajectory passes over it, producing the short term anomaly 
observed around $t_1$ in the light curve.

Although the two solutions produce similar anomalies, the position of the planet and the 
structure of the caustic differ between them.  In solution A, the planet generates a single 
four-fold caustic located on the same side as the planet relative to $M_1$. Given the 
relatively wide separation between the planet and the binary, this configuration suggests 
that the planet is most likely in a circumbinary orbit. In contrast, solution B yields two 
three-fold caustics situated on the opposite side of $M_1$. Because the planet in this case 
lies at approximately comparable projected separations from both stellar components, it is 
unlikely to be gravitationally bound to only one star (that is, a circumstellar configuration). 
Instead, this geometry is more plausibly interpreted as that of a distant planet whose apparent 
proximity to the binary results from projection effects.

In addition to the geometric differences, the two solutions yield markedly different mass 
ratios. For solution A, the estimated mass ratio is $q_3 \sim 0.55 \times 10^{-3}$, which 
is nearly 23 times smaller than that of solution B ($q_3 \sim 12.6 \times 10^{-3}$).

\begin{figure}[t]
\includegraphics[width=\columnwidth]{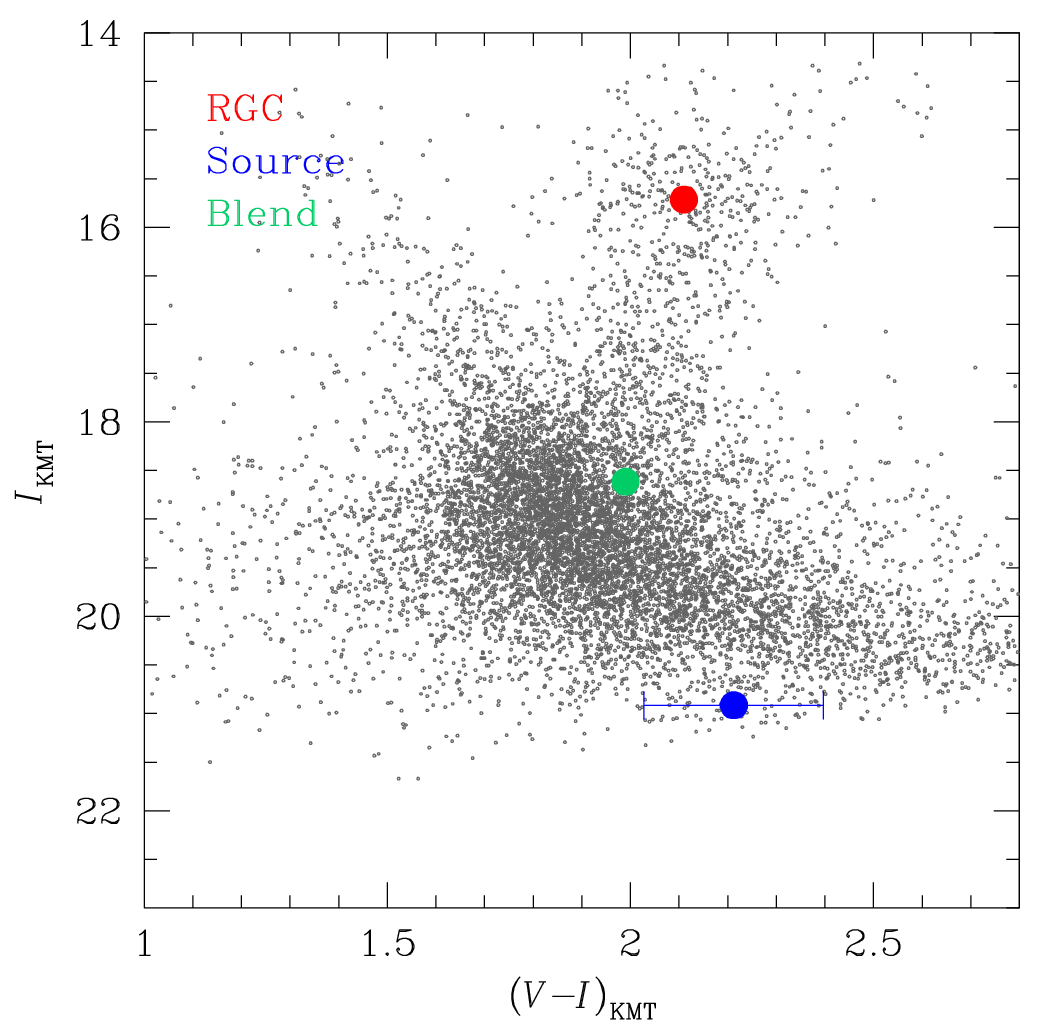}
\caption{
Location of the lensing source in the instrumental color-magnitude diagram. The
positions of the red giant clump (RGC) centroid and the blend are also indicated.
}
\label{fig:seven}
\end{figure}

\section{Angular Einstein radius} \label{sec:five}
   
For KMT-2016-BLG-1337, the normalized source radius $\rho$ was measured because the caustic
feature at epoch $t_3$ was well resolved. Combining this measurement with the angular radius 
of the source star ($\theta_*$) enables the determination of the angular Einstein radius as
\begin{equation}
\theta_{\rm E} = { \theta_* \over \rho}.
\label{eq1}
\end{equation}

We estimated the angular source radius by determining its reddening-corrected color and 
magnitude. To do this, we first derived the instrumental (uncalibrated) source magnitudes 
in the $I$ and $V$ bands by fitting the observed light curve to the model.  The source 
position was then placed on the instrumental color--magnitude diagram (CMD) for stars 
lying near the source.  In the next step, we calibrated the instrumental source color 
and magnitude using a reference point on the CMD. Specifically, we adopted the centroid 
of the red giant clump (RGC) as the reference because its de-reddened color and magnitude 
are well established from previous studies by \citet{Bensby2013} and \citet{Nataf2013}. 
For both the source flux measurement and the construction of the CMD, we utilized the pyDIA 
photometry code \citep{Albrow2017}. This code is designed to recover both the differential 
and reference fluxes from the template image, thereby enabling the reconstruction of the 
total source flux in instrumental units for each passband.\footnote{In contrast, the pySIS 
photometry code, on which the KMTNet photometric pipeline is based, measures only the 
differential flux with respect to the reference image.}

\begin{deluxetable}{lllllll}
\tablewidth{0pt}
\tablecaption{Source parameters, angular Einstein radius, and relative lens-source proper motion. \label{table:two}}
\tablehead{
\multicolumn{1}{c}{Parameter}     &
\multicolumn{1}{c}{Value    }     
}
\startdata
 $(V-I, I)$                  &   $(2.213 \pm 0.185, 20.914 \pm 0.025)$   \\
 $(V-I, I)_{\rm RGC}$        &   $(2.110, 15.714)                    $   \\
 $(V-I, I)_{{\rm RGC},0}$    &   $(1.060, 14.606)                    $   \\
 $(V-I, I)_0$                &   $(1.163 \pm 0.189, 19.806 \pm 0.032)$   \\
  Spectral type              &    K4V                                    \\
 $\theta_*$ ($\mu$as)        &   $0.57 \pm 0.12                      $   \\
 $\thetae$ (mas)             &   $0.53 \pm 0.11                      $   \\
 $\mu$ (mas/yr)              &   $4.73 \pm 0.98                      $   \\
\enddata
\end{deluxetable}

Figure~\ref{fig:seven} shows the locations of the source and RGC centroid in the instrumental 
CMD.  Using the offsets in color and magnitude, $\Delta (V-I, I)$, between the source and 
the RGC centroid, $(V-I, I)_{\rm RGC}$, the de-reddened source color and magnitude, $(V-I, 
I)_0$, were derived as
\begin{equation}
(V-I, I)_0 = (V-I, I)_{{\rm RGC},0} + \Delta(V-I, I),
\label{eq2}
\end{equation}
\hskip-5pt
where $(V-I, I)_{{\rm RGC},0}$ denote the de-reddened color and magnitude of the RGC 
centroid.  Table~\ref{table:two} lists the corresponding values of $(V-I, I)$, $(V-I, 
I)_{\rm RGC}$, $(V-I, I)_{{\rm RGC},0}$, and $(V-I, I)_0$. Based on the derived color 
and magnitude, the source is classified as a mid-K--type main-sequence star located in 
the Galactic bulge.

Based on the measured source color and magnitude, we derived the angular radius of the 
source, using the empirical color--surface brightness relation of \citet{Kervella2004}, 
which relates $(V-K, V)$ to the stellar angular size. To apply this relation, the observed 
$(V-I)$ color was converted to $(V-K)$ using the color--color calibration of 
\citet{Bessell1988}. With the resulting estimate of $\theta_*$, we then calculated the 
angular Einstein radius, $\theta_{\rm E}$, from Equation~(\ref{eq1}). Combining 
$\theta_{\rm E}$ with the measured event timescale, $t_{\rm E}$, gives the relative 
lens--source proper motion, $\mu = \theta_{\rm E} / t_{\rm E}$. The derived values of 
$\theta_*$, $\theta_{\rm E}$, and $\mu$ are summarized in Table~\ref{table:two}.

\begin{figure}[t]
\includegraphics[width=\columnwidth]{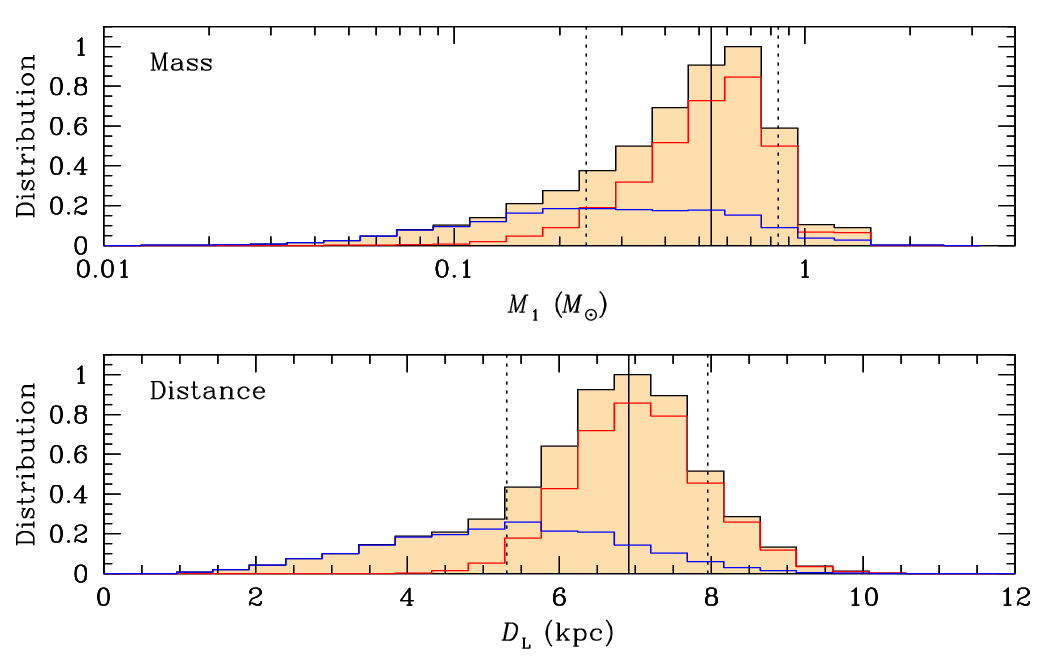}
\caption{
Bayesian posterior distributions for the mass of the primary (heaviest) lens component
($M_1$) and the distasixnce to the planetary system.  In each panel, the blue and red 
curves represent the contributions from the disk and bulge lens populations, respectively, 
while the black curve shows the combined posterior distribution. The solid vertical line 
marks the median value, and the two dotted lines indicate the 1$\sigma$ confidence interval.
}
\label{fig:eight}
\end{figure}

\section{Nature of the planetary system} \label{sec:six}

We characterize the planetary system by determining the component masses and the distance 
to the lens. These parameters were derived through a Bayesian analysis of the event. The 
details of this procedure have been described in numerous previous studies and are not 
repeated here. In brief, the analysis combines the measured microlensing observables, 
namely the event timescale ($t_{\rm E}$) and the angular Einstein radius ($\theta_{\rm E}$), 
with Galactic priors that describe the spatial, kinematic, and mass distributions of potential 
lenses. A large ensemble of synthetic microlensing events is generated from these priors, and 
each simulated event is assigned a statistical weight according to its consistency with the 
measured lensing observables.  The resulting weighted ensemble yields posterior probability 
distributions for the lens mass ($M$) and distance ($D_{\rm L}$).  For further details of the 
analysis procedure, we refer the reader to \citet{Han2025a}.

\begin{deluxetable}{lllllll}
\tablewidth{0pt}
\tablecaption{Physical parameters of the planetary system. \label{table:three}}
\tablehead{
\multicolumn{1}{c}{Parameter }     &
\multicolumn{1}{c}{Solution A}     &
\multicolumn{1}{c}{Solution B}     
}
\startdata
  $M_1$ ($M_\odot$)    &  $0.54^{+0.30}_{-0.30}$    &  $\leftarrow          $  \\   [0.8ex]
  $M_2$ ($M_\odot$)    &  $0.40^{+0.22}_{-0.22}$    &  $0.44^{+0.24}_{-0.25}$  \\   [0.8ex]
  $M_3$ ($M_{\rm J}$)  &  $0.28^{+0.15}_{-0.16}$    &  $7.11^{+3.93}_{-3.98}$  \\   [0.8ex]
  $\dl$ (kpc)          &  $6.92^{+1.04}_{-1.61}$    &  $6.95^{+1.04}_{-1.61}$  \\   [0.8ex]
  $a_{\perp,2}$ (au)   &  $3.54^{+0.53}_{-0.82}$    &  $3.65^{+0.55}_{-0.85}$  \\   [0.8ex]
  $a_{\perp,3}$ (au)   &  $3.97^{+0.60}_{-0.92}$    &  $1.49^{+0.22}_{-0.35}$  \\   [0.8ex]
  $p_{\rm disk}$       &  $34\%                $    &  $\leftarrow          $  \\   [0.8ex]
  $p_{\rm bulge}$      &  $66\%                $    &  $\leftarrow          $  \\   [0.8ex]
\enddata
\end{deluxetable}

Figure~\ref{fig:eight} presents the posterior distributions for the mass of the most 
massive lens component, $M_1$, and the distance to the lens system derived from the 
Bayesian analysis.  Table~\ref{table:three} summarizes the physical properties of the 
planetary system, including the individual masses of the lens components ($M_1$, $M_2$, 
and $M_3$), the distance to the system ($D_{\rm L}$), and the projected separations of 
$M_2$ and $M_3$ from $M_1$ ($a_{\perp,2}$ and $a_{\perp,3}$).  For each parameter, the 
reported value corresponds to the median of the posterior distribution, while the lower 
and upper uncertainties are defined by the 16th and 84th percentiles, respectively. The 
tables also list the probabilities that the lens resides in the Galactic disk ($p_{\rm disk}$) 
or in the bulge ($p_{\rm bulge}$).

For both solutions, the physical parameters of the host binary stars are essentially 
identical. The host system comprises two early M-type dwarf stars with masses of 
approximately $M_1 \sim 0.54~M_\odot$ and $M_2 \sim 0.40~M_\odot$, separated by a 
projected distance of about $a_{\perp,2} \sim 3.5$~au. The binary system is located at 
a distance of $\dl \sim 7$~kpc from the Sun in the direction of the Galactic center. The 
Bayesian analysis further indicates that the system is roughly twice as likely to reside 
in the Galactic bulge as in the disk.

In contrast, the physical properties of the planet inferred from solutions A and B differ
substantially. For solution A, the estimated planet mass is $\sim 0.3~M_{\rm J}$, comparable 
to that of Saturn in the Solar System. In solution B, however, the planet is much more massive, 
with a derived mass of $\sim 7.1~M_{\rm J}$. The inferred locations of the planet relative 
to the host binary stars also differ markedly between the two solutions. As illustrated in 
Figure~\ref{fig:six}, in solution A the planet lies on one side of the host binary, whereas 
in solution B it is positioned between the two stellar components. The projected separations 
from the primary star are $a_{\perp,3} \sim 4.0$~au for solution A and $a_{\perp,3} \sim 1.5$ 
au for solution B.

\section{Summary and conclusion}  \label{sec:seven}
   
We have analyzed the microlensing event KMT-2016-BLG-1337, which reveals the presence of 
a planetary companion in a binary system of low-mass stars.  The event was observed by 
the KMTNet survey and independently confirmed by the MOA and OGLE surveys.  The light 
curve exhibits a short-term anomaly superposed on a typical binary-lens profile. Modeling 
of the event shows that a triple-lens single-source configuration provides the best 
description, implying that the lens system consists of a planet orbiting a binary stellar 
host.

 The analysis yields two viable solutions that reproduce the observed anomaly equally well. 
In solution A, the planet has a mass of $M_3 \sim 0.3~M_{\rm J}$ at a projected separation 
of $a_{\perp,3} \sim 4~{\rm au}$, while in solution B, the planet is more massive ($M_3 \sim
7~M_{\rm J}$) and located closer to the host stars ($a_{\perp,3} \sim 1.5~{\rm au}$). Although
the two models differ in their interpretation of the planetary anomaly, the available data do 
not allow the degeneracy to be conclusively resolved, primarily due to missing coverage during 
the caustic entrance phase. A Bayesian analysis based on Galactic priors indicates that the 
host system is composed of two early M-dwarf stars with masses of $M_1 \sim 0.54~M_\odot$ and 
$M_2 \sim 0.40~M_\odot$, separated by $a_{\perp,2} \sim 3.5~{\rm au}$ and located at $\dl \sim 
7~{\rm kpc}$ toward the Galactic bulge.

The event KMT-2016-BLG-1337L underscores the capability of microlensing to reveal planets 
in dynamically complex stellar environments, including systems that are inaccessible to 
conventional detection techniques. This expands the census of planets in multiple-star 
systems and contributes to a more comprehensive understanding of planet formation in such 
environments.

\begin{acknowledgements}
C.H. was supported by the Chungbuk National University 2025 NUDP program and the National 
Research Foundation of Korea (RS-2025-21073000).
This research has made use of the KMTNet system operated by the Korea Astronomy and Space Science 
Institute (KASI) at three host sites of CTIO in Chile, SAAO in South Africa, and SSO in Australia. 
Data transfer from the host site to KASI was supported by the Korea Research Environment Open NETwork 
(KREONET). 
The OGLE project has received funding from the Polish National Science
Centre grant OPUS-28 2024/55/B/ST9/00447 to AU.
H.Y. and W.Z. acknowledge support by the National Natural Science Foundation of China 
(Grant No. 12133005). 
The MOA project is supported by JSPS KAKENHI Grant Number 
JP16H06287, JP22H00153 and 23KK0060.
C.R. was supported by the Research fellowship of the Alexander von Humboldt Foundation.
\end{acknowledgements}



\bibliographystyle{aasjournal}
\bibliography{pasp_refs}

\end{document}